# ATHENA X-IFU Demonstration Model: First joint operation of the main TES Array and its Cryogenic AntiCoincidence Detector (CryoAC)

M. D'Andrea[*,1] • K. Ravensberg[5] • A. Argan[1] • D. Brienza[1] • S. Lotti[1] • C. Macculi[1] • G. Minervini[1] • L. Piro[1] • G. Torrioli[2] • F. Chiarello[2] • L. Ferrari Barusso[3] • M. Biasotti[3] • G. Gallucci[3,4] • F. Gatti[3] • M. Rigano[3] • H. Akamatsu[5] • J. Dercksen[5] • L. Gottardi[5] • F. de Groote[5] • R. den Hartog[5] • J.-W. den Herder[5] • R. Hoogeveen[5] • B. Jackson[6] • A. McCalden[5] • S. Rosman[5] • E. Taralli[5] • D. Vaccaro[5] • M. de Wit[5] • J. Chervenak[7] • S. Smith[7] • N. Wakeham[8,7]

[*]*Now at INAF/IAPS, previously working at SRON*
[1]*INAF/IAPS Roma, Via del Fosso del Cavaliere 100, 00133 Roma, Italy*
[2]*CNR/IFN Roma, Via Cineto Romano 42, 00156 Roma, Italy*
[3]*University of Genoa, Via Dodecaneso 33, 16146 Genova, Italy*
[4]*INFN Genova, Via Dodecaneso 33, 16146 Genova, Italy*
[5]*SRON Leiden, Niels Bohrweg 4, 2333 CA Leiden, The Netherlands*
[6]*SRON Groningen, Landleven 12, 9747 AD Groningen, The Netherlands*
[7]*NASA GSFC, Greenbelt, MD 20771, USA*
[8]*CSST, University of Maryland, Baltimore, MD 21250, USA*

**Abstract** The X-IFU is the cryogenic spectrometer onboard the future ATHENA X-ray observatory. It is based on a large array of TES microcalorimeters, which works in combination with a Cryogenic AntiCoincidence detector (CryoAC). This is necessary to reduce the particle background level thus enabling part of the mission science goals. Here we present the first joint test of X-IFU TES array and CryoAC Demonstration Models, performed in a FDM setup. We show that it is possible to operate properly both detectors, and we provide a preliminary demonstration of the anti-coincidence capability of the system achieved by the simultaneous detection of cosmic muons.

**Keywords** X-rays: detectors • ATHENA X-IFU • TES • Anticoincidence

# 1 Introduction

ATHENA [1] is a Large X-ray observatory (launch by ESA in 2030s), aimed at studying the Hot and Energetic Universe [2]. The X-ray Integral Field Unit (X-IFU [3]) is its cryogenic spectrometer, able to perform simultaneously detailed imaging and high-energy resolution spectroscopy ($\Delta E_{FWHM}$ < 2.5 eV @ 7 keV). The core of the instrument is an array of ~3000 Transition Edge Sensor (TES) microcalorimeters. The TES array alone is not able to distinguish between target X-ray photons and background particles depositing energy in the detector band, seriously limiting the instrument sensitivity. This particle background is constituted by a primary component of both solar and Galactic Cosmic Rays origin, and by secondary particles generated inside the spacecraft. To deal with this issue, the Focal Plane Assembly (FPA) hosts a Cryogenic AntiCoincidence detector (CryoAC [4]). It is a TES-based detector, placed < 1 mm underneath the TES array. While X-ray photons are absorbed in the TES array, background particles deposit energy in both detectors, producing a coincidence signal that allows vetoing these unwanted events (Fig. 1). The CryoAC allows to reduce the X-IFU particle background by a factor ~50, reaching the scientific requirement of the mission. A detailed review of the X-IFU particle background issue can be found in Ref. [5].

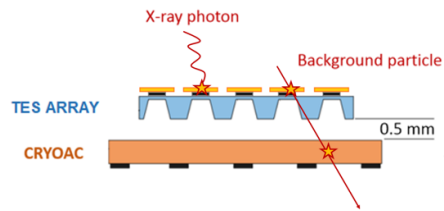

**Fig. 1** : Schematics of the working principle of the CryoAC combined with TES array. X-ray photons are absorbed by the TES array, whereas background particles deposit energy on both detectors, producing a coincidence signal. (Color figure online)

Ensuring mechanical, thermal and electromagnetic compatibility between TES array and CryoAC is a challenge in the FPA development. In the context of the development of the X-IFU Demonstration Model (DM), we performed the first joint test of TES array and CryoAC DMs. The main goals of the activity have been to demonstrate the proper simultaneous operation of the two detectors, to evaluate the crosstalk between them and to give a preliminary demonstration of the anticoincidence capability of the system.

# First joint test between ATHENA X-IFU TES Array and CryoAC

## 2 Experimental Setup

The test has been performed in the *SRON FDM 40-pixel-B setup*. The core of the setup is the detector plate, in which TES array and CryoAC DMs chips are mounted at 0.5 mm distance, in a relative position representative of the one in the X-IFU FPA (Fig. 2).

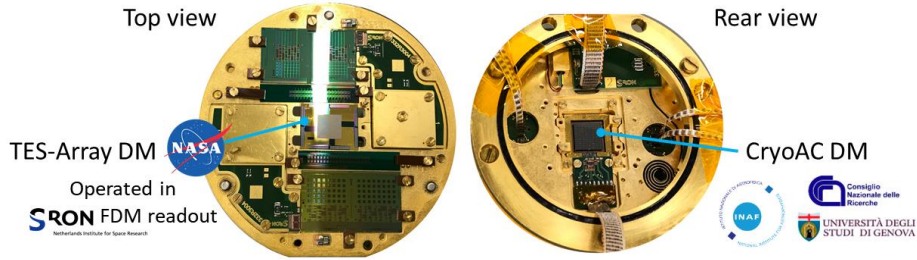

**Fig. 2** The detectors plate inside the SRON 40-pixel-B setup, showing both TES array DM (Top view) and CryoAC DM (Rear view) (Color figure online).

The experiment has been integrated on the mixing chamber of a dilution refrigerator, suspended by Kevlar strands (to damp microvibrations induced by pulse tube operation) and shielded at cold by a superconducting niobium shield. The setup hosted a $^{55}$Fe source to test the TES array response at 6 keV, an active thermal control system (typical thermal stability ~ 2 $\mu K_{RMS}$ at 50 mK) and a magnetic coil to apply a magnetic field perpendicular to the TES array to optimize the working point.

### 2.1 TES Array Demonstration Model

The TES array DM is a uniform 32x32 pixels array fabricated at NASA/GSFC. The main characteristics of the chip are summarized in Table 1.

**Table 1** TES Array Demonstration Model chip (ATH-1 G) characteristics

| Component | Characteristics |
|---|---|
| TESs | Mo/Au bilayer (35 nm/108 nm thick), 100x100 $\mu m^2$, $T_C$ ~ 87mK, $R_N$ ~ 51 m$\Omega$ |
| Absorbers | Au/Bi (2.50 $\mu m$/3.39 $\mu m$ thick), 240x240 $\mu m^2$, 250 $\mu m$ pitch |
| Membranes | SiN (0.5 $\mu m$ thick) |
| Leads | Nb (bottom)/SiO2/Nb (top) (154 nm/260 nm/212 nm thick), width: 6 $\mu m$ (bottom)/3 $\mu m$ (top) |

For details about TES array fabrication and design refer to [6]. The detector has been operated in the Frequency Domain Multiplexing (FDM) readout developed at SRON, which was the baseline readout method for the X-IFU when this experiment was planned. Here, we connected simultaneously 20 pixels of the array. A full description of the FDM readout scheme can be found in [7].

2.2 CryoAC Demonstration Model

The CryoAC DM is a single pixel detector, based on a large area (1cm$^2$) Silicon absorber. This is sensed by a network of 96 Ir/Au TESs connected in parallel, and readout by a SQUID operated in the standard Flux Locked Loop (FLL) configuration. The TES network is designed to achieve efficient athermal phonon collection, and it features anti-inductive niobium wirings to limit the electromagnetic coupling with the TES array. Platinum heaters are deposited on the absorber for calibration and diagnostic purposes. Details about the CryoAC DM are in ref. [8] (fabrication) and [9] (characterization and test).

**3 Compatibility Test**

In this section, we report the main results of the measurements performed to verify the proper simultaneous operation of the two detectors.

3.1 Impact of the CryoAC operation on TES array performance

First, we verified that TES array pixels show similar performance before and after the introduction of the CryoAC in the setup. In Fig. 3 - Left are shown the performance (i.e. $\Delta E_{FWHM}$ measured on Mn-k$\alpha$ complex at 6 keV) of 4 reference pixels operated at different FDM bias frequencies, from 1 MHz to 4 MHz, measured in the 40-pixel-B setup with and without the CryoAC. No significant degradation has been observed in the two configurations, within a statistical accuracy around 0.3 eV. A more sensitive test will be presented in Sect. 3.3, with the dynamic crosstalk measurement.

3.2 CryoAC DM requirements verification

On the CryoAC side, we have verified the detector compliance with its functional requirements: low-energy threshold < 20 keV; operation at a thermal bath temperature $T_B$ = 50 mK; power dissipation at cold < 40 nW [10]. In the integrated setup the detector has been operated at $T_B$ = 50 mK, with an assessed total power dissipation $P_{TOT}$ = 3.3 nW and a trigger

**First joint test between ATHENA X-IFU TES Array and CryoAC**

threshold $E_{THR}$ = 8 keV, fully fulfilling the DM requirements. Fig. 3 - Right shows the typical pulse detected by the CryoAC after the interaction with a cosmic muon.

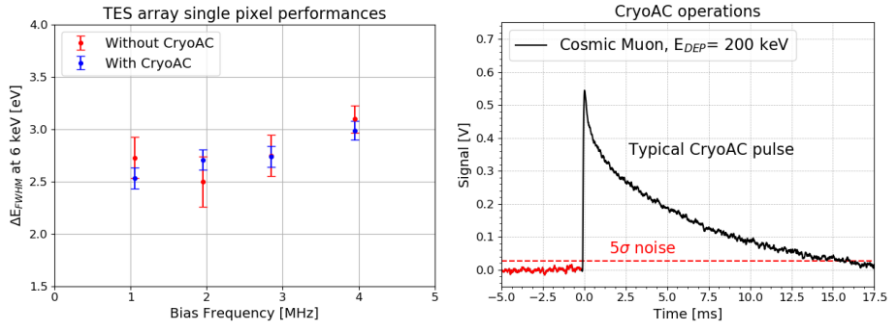

**Fig. 3** TES array and CryoAC DMs operations. *Left:* TES array single pixel performances before and after the integration of the CryoAC in the experimental setup. *Right:* Typical pulse generated by a cosmic muon interacting with the CryoAC. (Color figure online).

3.3 Magnetic coupling and crosstalk evaluation

To assess the impact of CryoAC operations on the magnetic environment at TES array level, we performed magnetic field scans on TES array pixels (Fig. 4). These measurements consist in varying the magnetic field normal to the TES array (via the setup magnetic coil), and monitoring its biased pixels baseline level. Typically, this level shows a maximum in correspondence of the optimal magnetic field value (i.e. the value that cancels the residual magnetic field at the pixel level).

Here, we performed the magnetic field scans while operating the CryoAC at different bias current (corresponding to different colors in Fig. 4). The scans have not highlighted any change in the magnetic environment at TES array level induced by the CryoAC, since the parabolic shape and the maximum position remain the same for each acquired curve. Differences in the absolute signal amplitude are due to slow thermal effects related to the change of the CryoAC power dissipation. We repeated this measurement for different pixels, obtaining similar results. We can conclude that the typical CryoAC DM bias currents do not have a significant magnetic impact on the TES array.

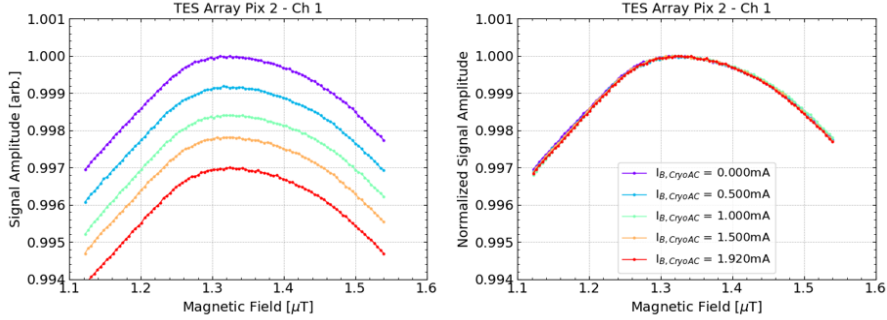

**Fig. 4** Magnetic scan performed on a TES array pixel for different CryoAC bias currents (different colors). *Left:* Baseline level of the pixel as a function of the magnetic field. *Right:* Same plot with the signal amplitude normalized at maximum level for each curve. (Color figure online).

We have also performed dynamic crosstalk measurements (Fig. 5). In this case, we have injected high-energy thermal pulses on the CryoAC via its onboard heater (energy > 1MeV, frequency: 10 Hz), and simultaneously looked at the signal on TES array blind pixels, using a trigger generated by the CryoAC via a dedicated custom electronics. Averaging the acquired signals, we have found no evidence of significant crosstalk on the TES array when the CryoAC develops a pulse, down to the level of 0.1 eV.

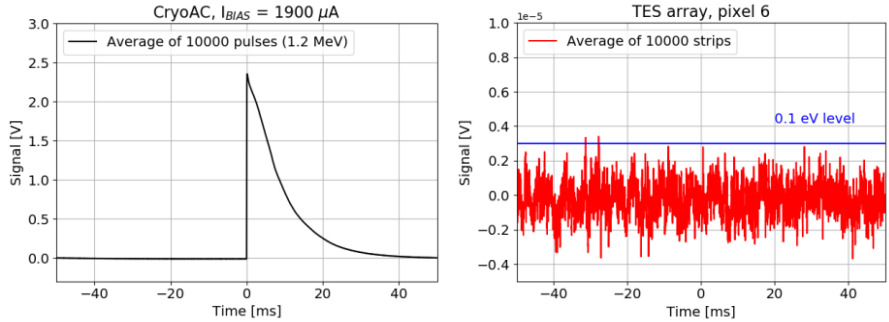

**Fig. 5** Dynamic crosstalk measurement. *Left:* Average of 10000 thermal pulses (1.2 MeV) generated in the CryoAC by the on-board heater. *Right:* Average of the 10000 simultaneous triggered strips of signal on a TES array pixel. (Color figure online)

The measurement has been repeated for different pixels and CryoAC bias currents, without noticing significant crosstalk effects.

# First joint test between ATHENA X-IFU TES Array and CryoAC

## 4 AntiCoincidence Measurements

Finally, we performed joint long measurements with both detectors to detect coincidence signals due to cosmic muons (Fig. 6). We operated simultaneously the CryoAC and 19 TES array pixels (multiplexing mode) for 890 ks (around 10 days), collecting 286 coincidence events. The observed count rate is 1.6 cts/cm$^2$/min, in agreement with the expectations for cosmic muons [11].

For all the collected events, we analyzed both the energy depositions on the CryoAC and on the TES array. The acquired spectra are shown in Fig. 7. The spectra are consistent with the expectations for Minimum Ionizing Particles (MIPs) [12], showing Landau distribution shapes and peaking energies of ~ 7 keV for the TES array and ~ 200 keV for the CryoAC.

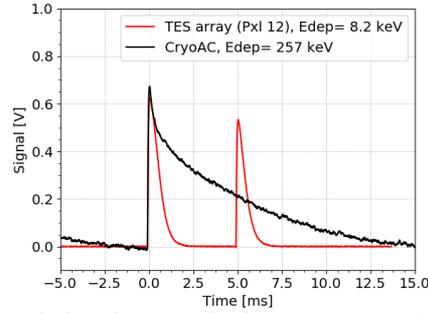

**Fig. 6** Pulses detected simultaneously by TES array (red line) and CryoAC (black line). On TES array, the first pulse is due to a background particle (detected also by the CryoAC), while the second one is due to an X-ray photon from $^{55}$Fe source (detected only by the TES array). (Color figure online)

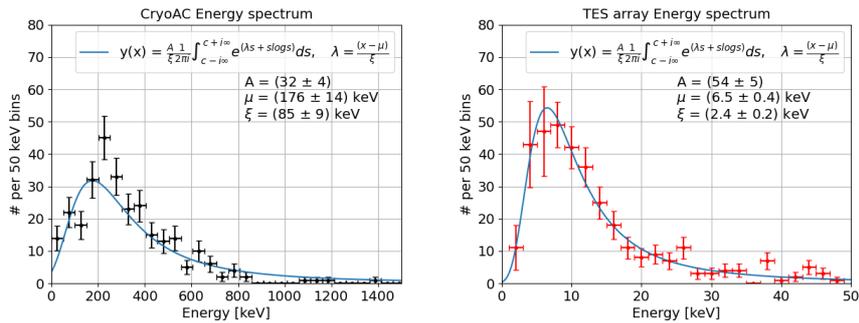

**Fig. 7** Energy depositions on CryoAC (*Left*) and TES array (*Right*) for detected coincidence events. The spectra have been fitted by Landau Distributions [13]. (Color figure online).

## 5 Conclusions

The first joint operation between the X-IFU TES array and CryoAC Demonstration Models has been performed in a FDM setup, showing that the detectors can properly operate together. No significant magnetic coupling between them has been detected, and no crosstalk has been measured down to the level of 0.1 eV on TES array pixels.

In the context of the X-IFU development, we shall note that the readout technology for the TES array has been changed close to the so far discussed integrated chipset test. The baseline moved from the FDM developed at SRON, used in this work, to the Time Division Multiplexing (TDM) scheme developed at NIST [6], where pixels are DC biased. Although the FDM has recently performed big jump ahead [7], TDM has been indeed considered by the system the most mature technology for X-IFU aims. A review about these technologies and the difference in the pixel optimization used in AC and DC biased multiplexing schemes can be found in [14].
About this work, we shall note that pixels operated under AC bias (FDM) suffer a lower magnetic field sensitivity than pixels operated under DC bias (TDM). Since the TES array and CryoAC interaction is strongly related to the magnetic coupling between the detectors, the compatibility measurements here presented shall be then integrated with a new set of measurements performed in a DC-bias setup.

The performed tests represents anyway an important step towards the FPA development. In particular, the simultaneous detection of cosmic muons on TES array and CryoAC has provided a first demonstration of the anticoincidence capabilities of the system, representing a milestone for the X-IFU project.

**Acknowledgements** This work has been supported by ASI (Italian Space Agency) through the Contracts no. 2015-046-R.0, 2018-11-HH.0 and 2019-27-HH.0, and by the ESA (European Space Agency) Core Technology Program (CTP) contract no. 4000114932/15/NL/BW. SRON Netherlands Institute for Space Research is supported financially by NWO, the Dutch Research Council, and this work is funded partly by NWO under the research programme Athena with project number 184.034.002. The material is also partly based upon work supported by NASA under award number 80GSFC21M0002.

# First joint test between ATHENA X-IFU TES Array and CryoAC